\documentclass[12pt]{article}
\usepackage{graphicx}
\begin{document}

\centerline{\bf \large Sociophysics Simulations I: Language Competition}

\bigskip
\centerline{Christian Schulze and Dietrich Stauffer}

\bigskip
\centerline{Institute for Theoretical Physics, Cologne University, D-50923 K\"oln, Euroland}

\begin{abstract}
Using a bit-string model similar to biological simulations, the competition 
between different languages is simulated both without and with spatial 
structure. We compare our agent-based work with differential equations and 
the competing bit-string model of Kosmidis et al.
\end{abstract}

\section{Introduction and Models}
According to the Bible, since the tower of Babylon was destroyed, humans speak
numerous (presently nearly $10^4$) different languages. Many of these face
extinction, and a few new ones may arise. Abrams and Strogatz \cite{strogatz}
described the competition between two languages by simple differential equations
averaging over all people, an approximation criticized in this Granada Seminar 
by Droz. Patriarca and Lepp\"anen \cite{finland}
applied these methods to a square lattice, with one language favoured in one 
half and the other language in  the other half. Other
models averaging over many people were published by Pagel \cite{pagel} and
Briscoe \cite{briscoe}.

Two agent-based models, with each person simulated separately, were proposed
independently in \cite{schulze,kosmidis}, using bit-strings as is customary 
for biological species \cite{eigen}. We shortly review here these agent-based
models; a longer review is given elsewhere \cite{stauffer}.


In \cite{schulze}, a language is given by a bit-string such that each different 
bit-string like 0011 and 1000 represents a different language; this model thus
simulates many languages. In the alternative model \cite{kosmidis} for only
two languages, the left part of the bit-string correponds to one and the right 
part to the other language; then a person with bit-string 0011 speaks the 
second language well, and the first not at all, while 1000 speaks the first
language badly, and the second not at all. We will later see how a 
re-interpretation gives good agreement between these models. Now we concentrate 
on the model of \cite{schulze}.

\section{Size and Lifetime}
Fig.1a shows the present histogram of the sizes of human languages, where the 
size is the number of people speaking mainly this language 
\cite{sutherland,science}. We see roughly a log-normal distribution, with an 
enhanced number at small sizes. Fig.1b shows a simulation made
as follows:

At each time step $t$ we determine first the fraction $x_i= N_i(t)/N(t)$ of 
people speaking language $i$. People flee from rare languages by selecting
with probability $1- x_i^2$ the language of one randomly selected person from 
the $N(t)$ survivors. (In \cite{schulze} a flight probability 
$(N(t)/N(t=\infty))(1-x_i)^2$ was simulated instead.) Then everybody dies with 
Verhulst probability $\propto N(t)$ due to overpopulation. The survivors 
produce one child each, which learns the language of the parent except that
with probability $p$ one of the $\ell$ bits in the bit-string is toggled. 
We see that the simulation of Fig.1b recovers the deviation from a log-normal 
distribution, which would be a parabola in these log-log plots. On the computer
we could simulate more languages than in reality but not as many people as
live now on Earth.
 
For low mutation probabilities $p < 1/4$, nearly everybody speaks one language  
(which in reality may correspond to the alphabet now in use), with a few mutants
of one bit only: Dominance. For higher mutation probabilities $> 1/4$, the 
population fragments into many different languages, whose sizes do not differ 
by orders of magnitude and may become equal if the total population size goes 
to infinity. Even in the stationary state, languages continuously face 
extinction and rebirth, and Fig.2 shows that most of these languages live only
one or a few iterations. Fig.2a holds for dominance and 8 bits, Fig.2b for
dominance or fragmentation at 8 or 16 bits, and Fig.2c for dominance at 16 bits
and populations between 1000 and ten million. 

We did not forget the fourth data set (fragmentation at 8 bits) in Fig.2b but 
there are no extinctions anymore in this case (we counted only extinctions for
$1000 < t < 2000$ to get rid of non-equilibrium effects at the beginning.) The
reason may be that for fragmentation we have many languages of roughly equal 
sizes, and if we wait long enough each of these languages eventually will 
die out accidentally. The larger the population is the longer we have to 
wait. And for 8 bits, when the one million simulated people are distributed
among only $2^8 = 256$ languages, the population is higher and the extinction 
time much higher than for the $2^{16} = 65536$ languages at 16 bits. 
Thus no extinctions were seen for 8 bits and fragmentation. On the 
dominance side at lower $p$, at each iteration many mutants are formed 
differing by one bit only from the preferred language and mutating back to it 
at as soon as possible; then we have data for both 8 and 16 bits. 

\section{Mixing} 
English has evolved as a mixture from French and German, and indeed starting
with two languages we did end up with one \cite{stauffer}. However, this was
one of the two original languages
and not a mixture. Kosmidis et al \cite{kosmidis}, on the other 
hand, also got mixture languages where in the simplest case on average each 
person takes half of the words from one and the other half from the other
language. In this aspect, their model \cite{kosmidis} at first seems better.
However, a re-interpretation of our model \cite{schulze} in their spirit
\cite{kosmidis} 
recovers the same mixture language.

For this purpose we follow \cite{kosmidis} and interpret the first half of
our bit-string as representing words from French, and the second half as
representing German words. Initially, half the population speaks French and
the other speaks German. In the dominance case, one of these two languages 
wins, and nearly everybody speaks it apart from minor variations: Wings in 
Fig.3. Nearly everybody has no bits set in one of the two 8-bit halves at $\ell
= 16$, and has all 8 bits set in the other half; exceptions are exponentially
rare as seen in Fig.3. For fragmentation instead of dominance, the parabolic 
curve in Fig.3 shows 
nearly identical results for the two halves: The most probable case is four
bits set in one half and four bits set in the other half, which is the desired 
mixture. 

Kosmidis et al \cite{kosmidis} also changed their many parameters such that
at the end nearly all bits were set: These are bilinguals speaking both
languages fluently.

Mixing of populations speaking different languages may also happen 
\cite{finland} if we
put the people onto a square lattice, where many people can live on each site.
Nevertheless they can move with probability 0.01 to a neighbouring site, and 
with probability $p$ accept the language of a randomly selected person living 
on the same site. Finally, during a mutation the bit is flipped randomly with 
probability $1-q$ while with probability $q$ it is taken from the corresponding
bit of another person. (This other person is selected randomly: with probability
1/2 from the same lattice, otherwise from one of the four neighbours.) Now 
a stable coexistence of two populations is possible, the French on one side
of Canal Street in New Orleans, and the English on the other side: Fig.4. With
an exponentially decaying probability someone lives as a language minority on
the wrong side of the street. The exponential decay constant, i.e. the 
interface width, seems to be independent of the lattice size, in contrast to 
surface roughening in the 2D Ising model. 
 
\section{Summary}

Not only our simulations on lattices like Fig.4 required agent -
based simulations, instead of differential equations averaging 
over the whole population. Also in the case without such spatial 
effects a differential approach for infinite populations would 
have been problematic: In the thermodynamic limit, loved by
physicists looking for analytic solutions and criticized 
elsewhere at this seminar for sociophysics, fragmentation would 
presumably have resulted in a delta function of the language size
distribution as opposed to Fig.1, and might have destroyed some
phase transitions of \cite{stauffer,schulze} 

The computer simulation of language competition, as opposed to the simulation
of learning a language by children or of the evolution of the first language out
of the sounds of a proto-language, is still in its infancy, both for mean
field approaches and for agent-based models. Everybody is invited to join.
 
We thank P.M.C. de Oliveira for the suggestion to simulate languages, and
L. Gallos for telling us at this seminar about \cite{kosmidis}.

\begin{figure}
\begin{center}
\includegraphics[angle=-90,scale=0.4]{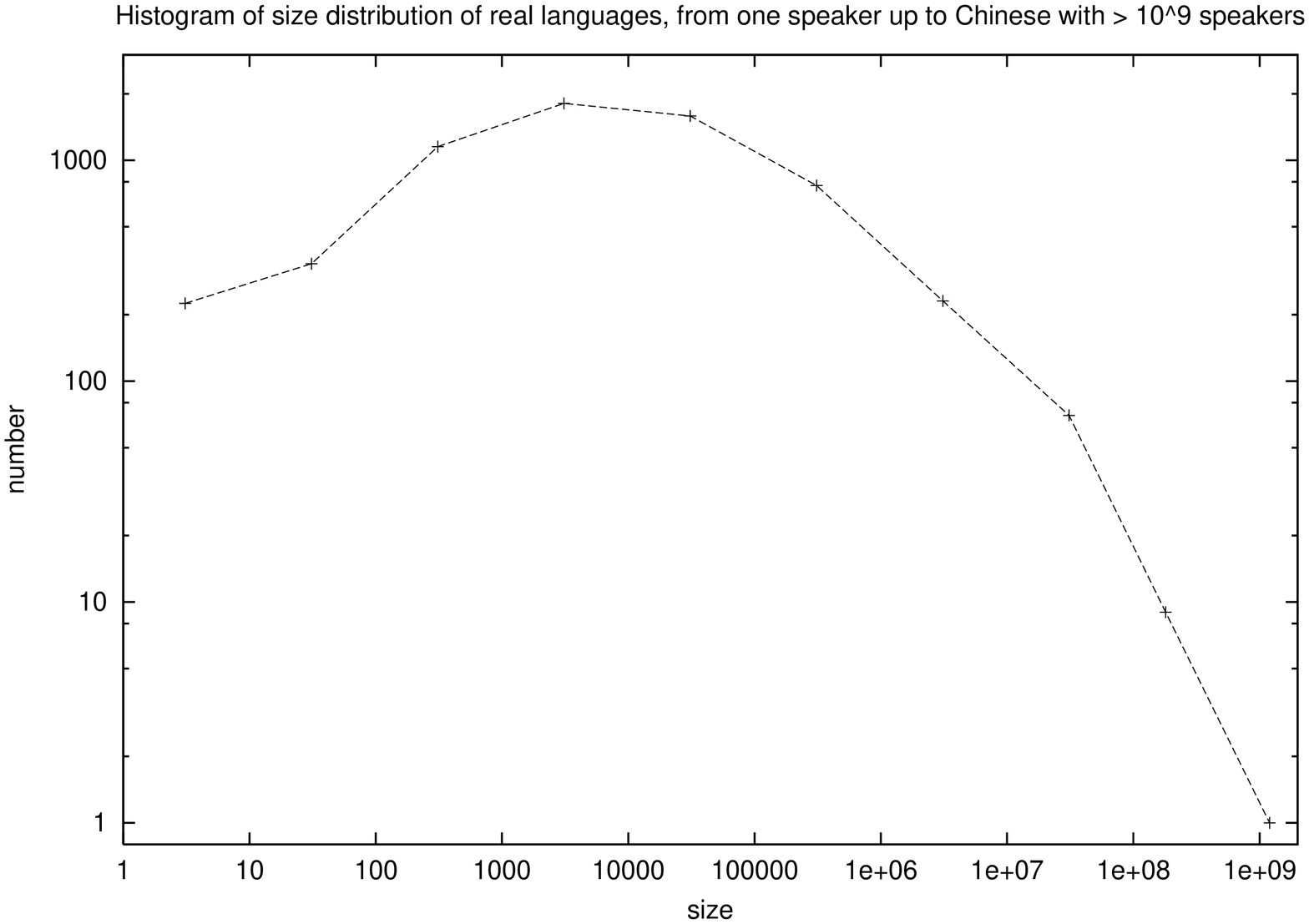}
\includegraphics[angle=-90,scale=0.4]{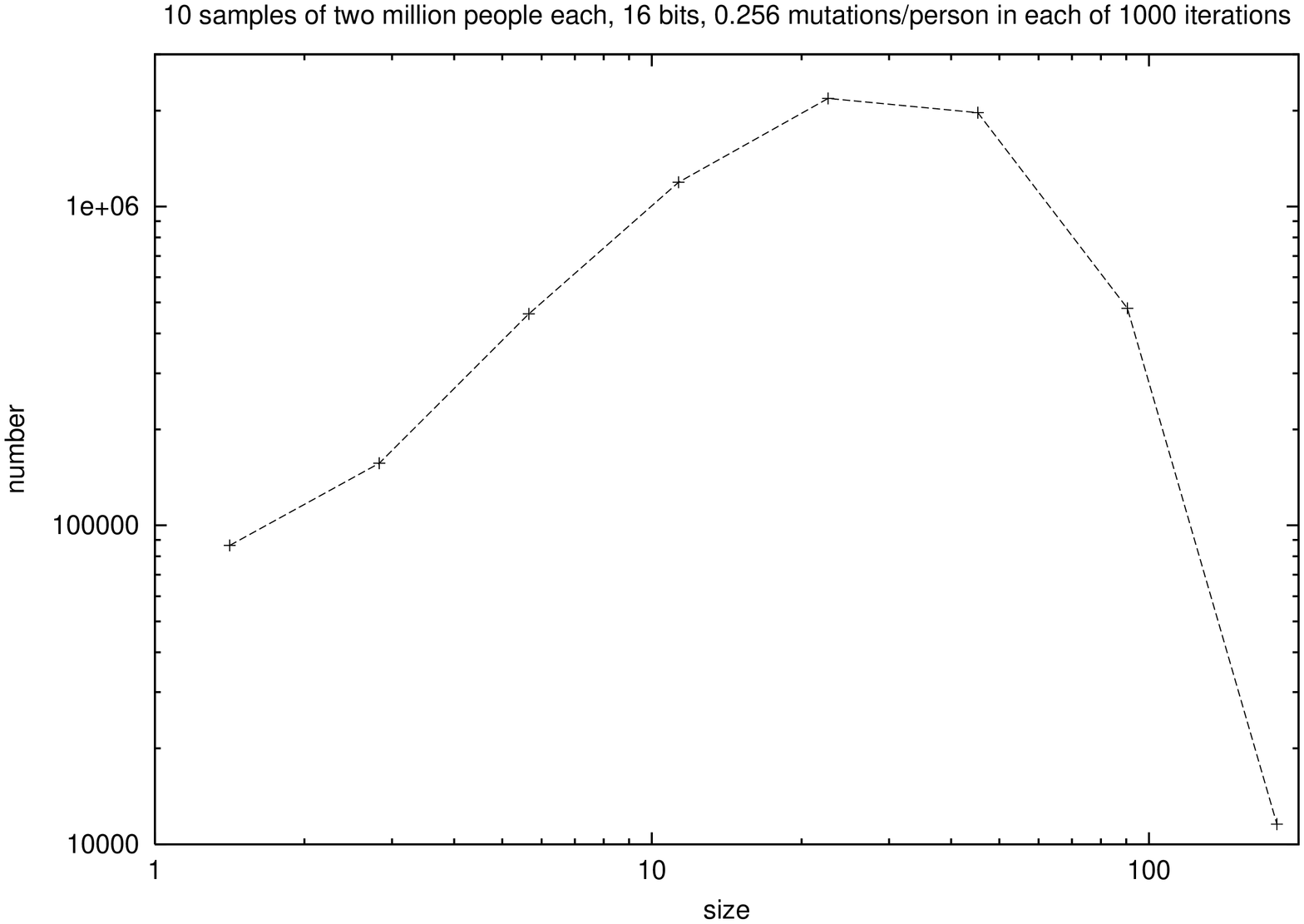}
\end{center}
\caption{Histogram of language sizes in reality (top, from \cite{stauffer}) and in our simulation 
(bottom; $p=0.16, \ell=16$).
}
\end{figure}

\begin{figure}
\begin{center}
\includegraphics[angle=-90,scale=0.30]{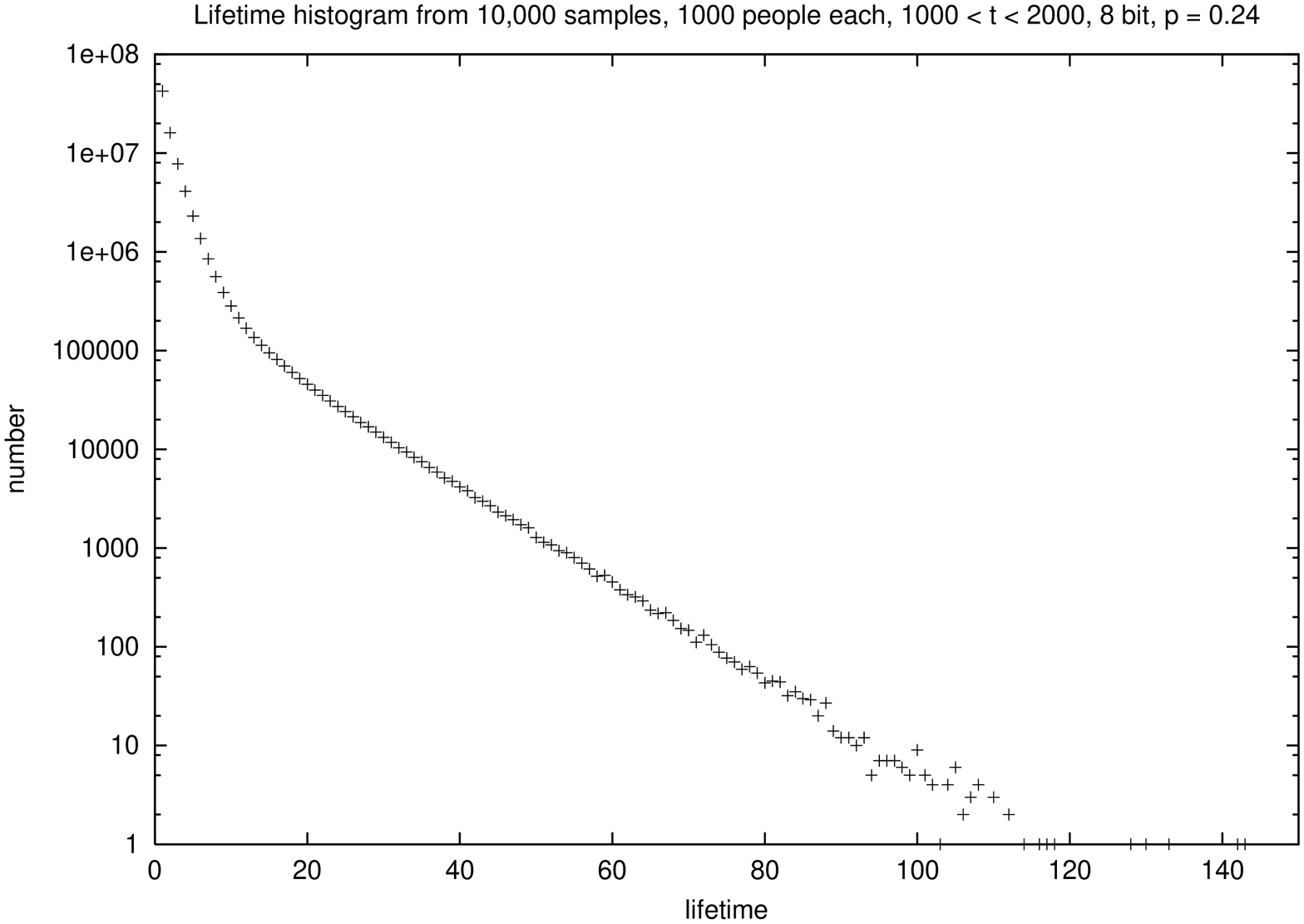}
\includegraphics[angle=-90,scale=0.30]{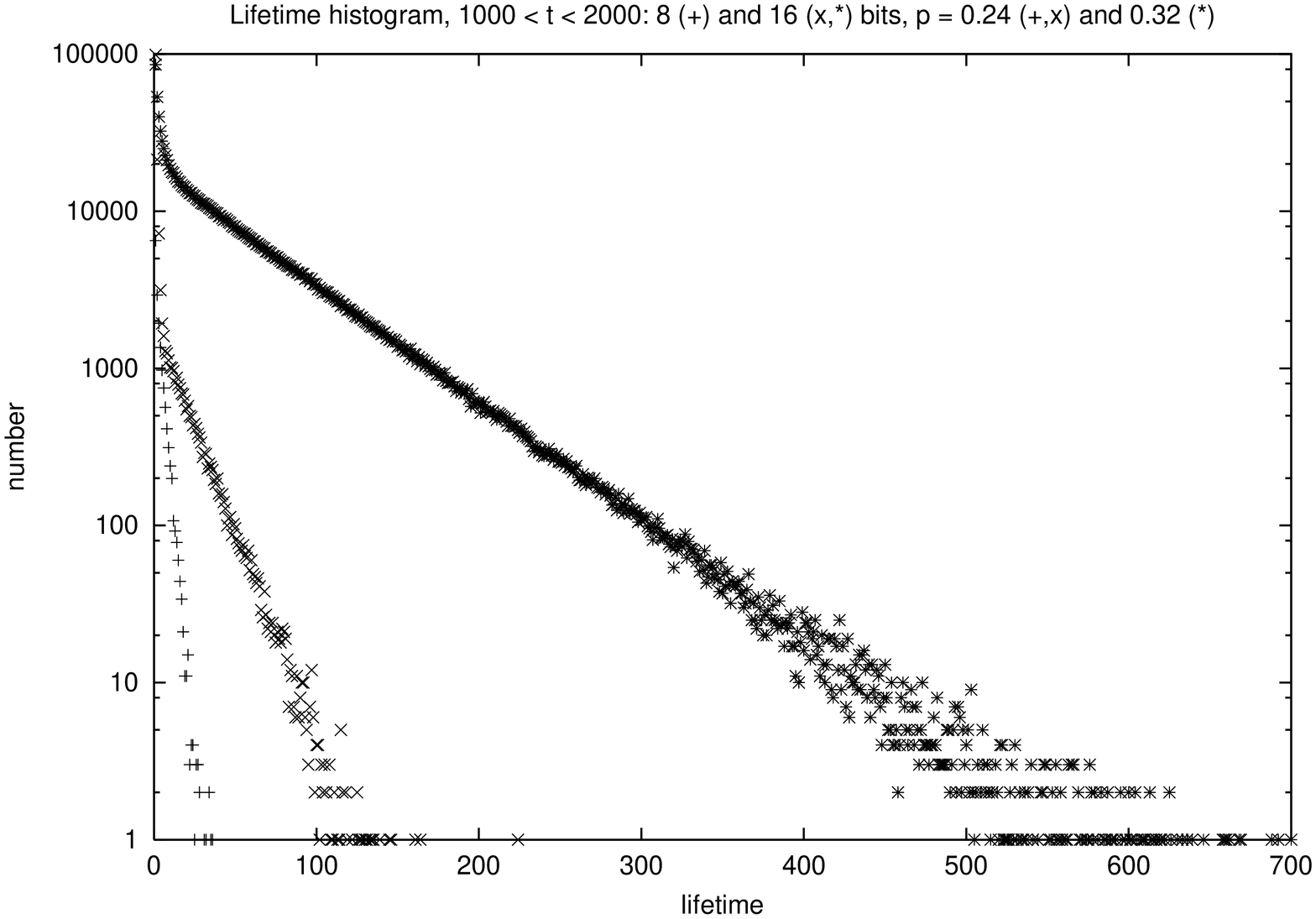}
\includegraphics[angle=-90,scale=0.30]{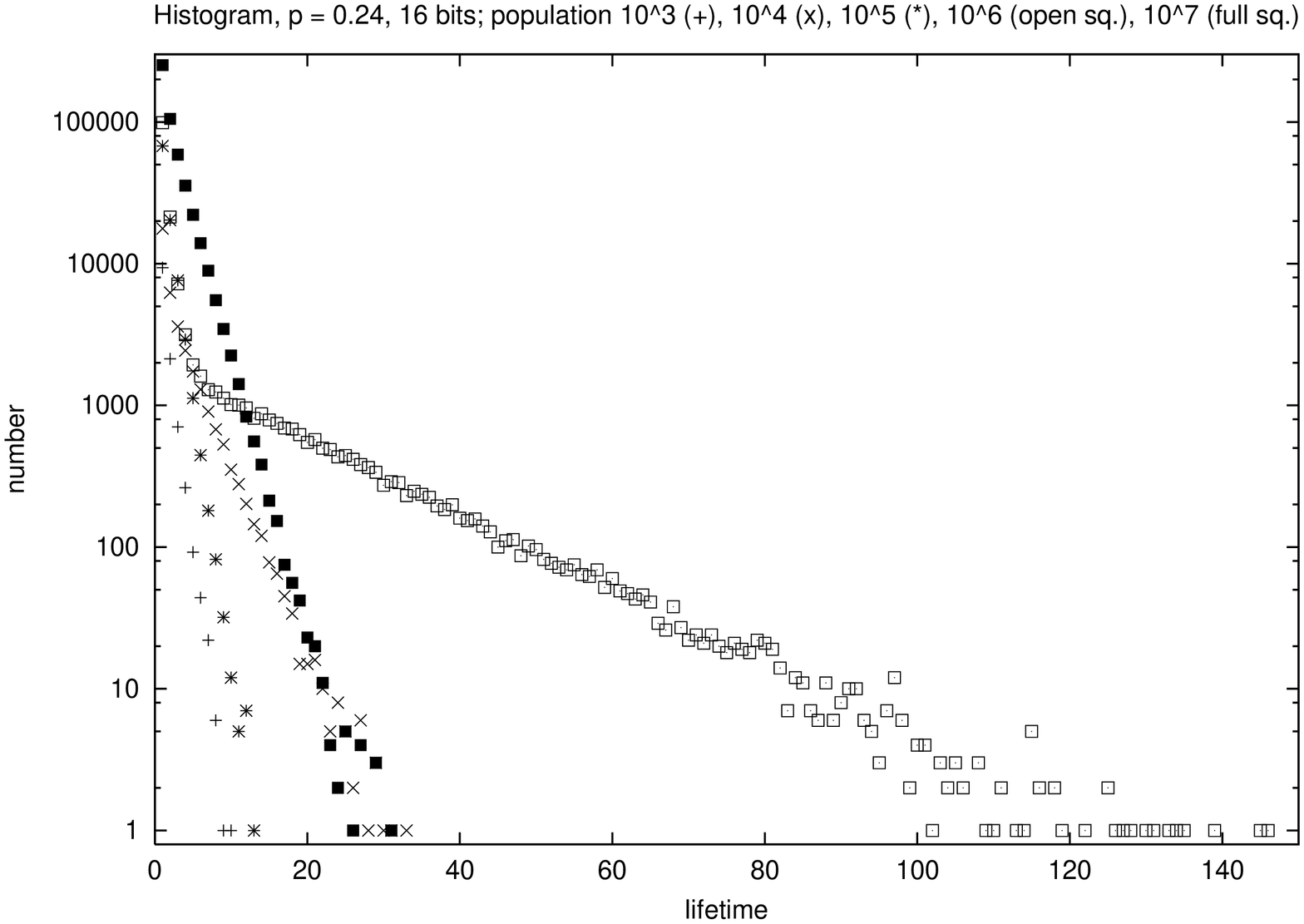}
\end{center}
\caption{Histogram of language lifetimes: dominance (top), dominance and 
fragmentation (center), dominance for different population sizes (bottom).
}
\end{figure}

\begin{figure}
\begin{center}
\includegraphics[angle=-90,scale=0.39]{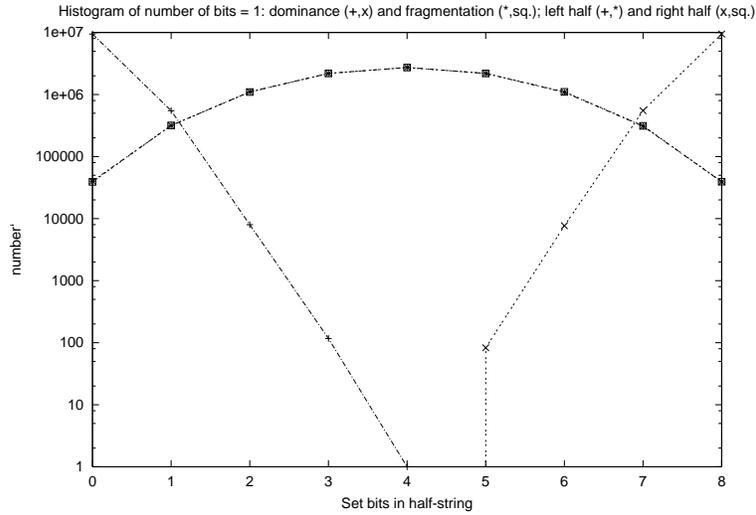}
\end{center}
\caption{Histogram of the number of bits set to one in the left and in the right
part of the bitstring with 16 bits. The parabola corresponds to fragmentation
(or mixing of two languages) while the two wings correspond to dominance of one
of the two initial languages.
}
\end{figure}

\begin{figure}
\begin{center}
\includegraphics[angle=-90,scale=0.39]{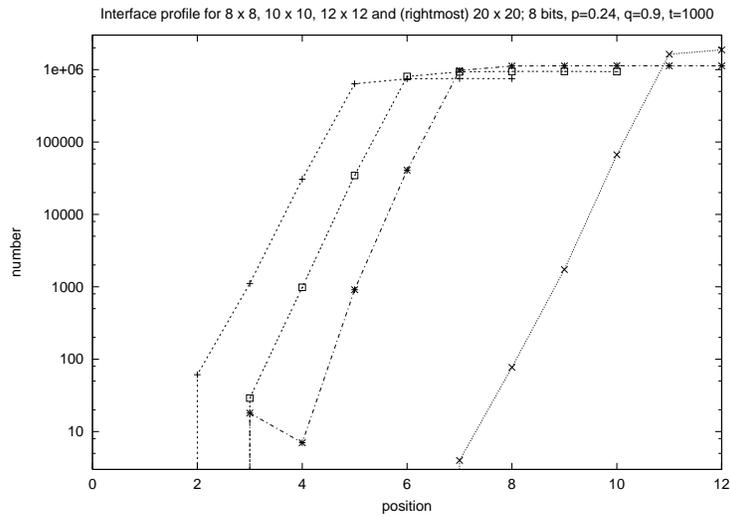}
\end{center}
\caption{Histogram of second language near the interface between two languages,
for $8 \times 8$ (left) to $20 \times 20$ (right) lattices, showing the 
exponential tunneling of the minority language into the space of the majority 
language.}
\end{figure}
\end{document}